\documentclass[12pt]{article}

\newcommand{\be}{\begin{equation}}
\newcommand{\ba}{\begin{eqnarray}}
\newcommand{\ea}{\end{eqnarray}}
\newcommand{\ee}{\end{equation}}

\newcommand{\z}{{\mathbf Z }}

\newcommand{\beq}{\begin{equation}}
\newcommand{\eeq}{\end{equation}}
\newcommand{\beqa}{\begin{eqnarray}}
\newcommand{\eeqa}{\end{eqnarray}}
\newcommand{\CR}{\nonumber \\}

\newcommand{\E}{\epsilon}

\newcommand{\unit}{\hbox to 3.8pt{\hskip1.3pt \vrule height 7.4pt
    width .4pt \hskip.7pt \vrule height 7.85pt width .4pt \kern-2.4pt
    \hrulefill \kern-3pt \raise 3.7pt\hbox{\char'40}}}

\def\matt[#1,#2,#3,#4]{\left(%
\begin{array}{cc} #1 & #2 \\ #3 & #4 \end{array} \right)}

\begin{document}
\begin{titlepage}
\thispagestyle{empty}
\begin{flushright}
YITP-08-54
\end{flushright}

\bigskip

\begin{center}
\noindent{\Large \textbf{
On M5-branes in ${\cal N}=6$ Membrane Action
}}\\
\vspace{2cm}
\noindent{

Seiji Terashima\\

{\it Yukawa Institute for Theoretical Physics, 
%\hspace{40mm}
Kyoto University, \\
Kyoto 606-8502, Japan}\\
%E-mail: \email{terasima@yukawa.kyoto-u.ac.jp}\\
}

\vskip 2em
\bigskip

\end{center}
\begin{abstract}

In this note we study M5-branes in
the multiple membrane action which is recently proposed 
by Aharony-Bergman-Jafferis-Maldacena.
We write down the ${\cal N}=6$ 
supersymmetry transformation of the action
and obtain 1/2 BPS equations and their solutions.
They are expected to represent membranes 
ending on a M5-brane.
We also consider the M5-M2 bound state
in the action.

\end{abstract}
\end{titlepage}

\newpage

%\tableofcontents

\section{Introduction}

Since an action of multiple M2-branes
proposed by the Bagger and Lambert \cite{BL2}
(see for earlier works \cite{BaHa, early}),
it has been studied intensively \cite{BLGr1}-\cite{BLGr2}. 
Recently, 
a three dimensional ${\cal N}=6$ supersymmetric Chern-Simons-matter 
conformal field theory
with gauge group $U(N) \times U(N)$ was proposed
as an action of the low energy limit of 
$N$ M2-branes on ${\bf C}^4 / \z_k$ 
by Aharony-Bergman-Jafferis-Maldacena (ABJM) \cite{ABJM}.
Many aspects of the theory have been studied 
\cite{Klebanov}-\cite{TeYa}.

The M5-branes are also interesting and 
still mysterious objects in M-theory. 
In this paper, we study the BPS equations 
of this ABJM action, which will describe 
the M5-brane.
We find solutions of these equations.
%for $N=2$.
These BPS equations are
analogues of the Basu-Harvey equation
\cite{BaHa} and
we expect that 
the solutions represent
$N$ M2-branes ending on the M5-brane.

We also expect that
the flat M5-branes will be constructed from 
infinitely many M2-branes, as the D4-D2 bound state.
This M5-M2 bound state 
has different supersymmetries from the ones 
which M5-branes have. 
Thus M5-M2 bound state on the orbifold will not be 
BPS and
we can not expect that there is the BPS solution corresponding to
this bound state in the ABJM action.
Therefore, instead of the BPS equation, 
we will discuss solutions of the equations of motion,
which will describe the M5-M2 bound state. 

The organization of this paper is as follows. 
In section two we briefly review the 
ABJM action and present an manifest ${\cal N}=6$
SUSY transformation of this action. 
In section three we 
study the BPS equations of the ABJM action
and their solutions.
The M5-M2 bound state is discussed in section four.
In section five we draw conclusions
and discuss future problems.

\section{${\cal N}=6$ SUSY action and SUSY transformation}

In this section we will briefly review 
the ABJM action.
The fields in the ABJM action are
$U(N) \times U(N)$ gauge fields
$A_\mu$ and $\hat{A}_\mu$,
four $U(N) \times U(N)$ bi-fundamental bosonic fields
$Y^A$ and fermionic spinor fields $\psi_A$, where $A=1,2,3,4$.

The $SU(4)$ invariant action of this theory is explicitly 
given by \cite{Klebanov, ABJM}
\begin{eqnarray}
S=\int d^3 x \left[ \frac{k}{4 \pi} \varepsilon^{\mu\nu\lambda} \mathrm{Tr} \left(
A_{\mu} \partial _{\nu} A_\lambda + \frac{2i}{3} A_{\mu} A_{\nu} A_{\lambda} 
- \hat{A}_{\mu} \partial_{\nu} \hat{A}_{\lambda} 
- \frac{2i}{3} \hat{A}_{\mu} \hat{A}_{\nu} \hat{A}_{\lambda} \right) \right. \CR
\left. - \mathrm{Tr} D_{\mu} Y_A^{\dagger} D^{\mu} Y^A 
- i \mathrm{Tr} \; \psi^{A \dagger} \gamma^{\mu} D_{\mu} \psi_A
 - V_{\mathrm{bos}} - V_{\mathrm{ferm}} \right]
\label{action}
\end{eqnarray}
with the potentials
\begin{eqnarray}
V_{bos}= -\frac{4 \pi^2 }{3 k^2}
\mathrm{Tr} \left( 
Y^A Y_A^\dagger Y^B Y_B^\dagger Y^C Y_C^\dagger
+ Y_A^\dagger Y^A Y_B^\dagger Y^B Y_C^\dagger Y^C \right. \CR
\left. 
+4 Y^A Y_B^\dagger Y^C Y_A^\dagger Y^B Y_C^\dagger
-6 Y^A Y_B^\dagger Y^B Y_A^\dagger Y^C Y_C^\dagger \right),
\end{eqnarray}
and
\begin{eqnarray}
V_{ferm}= -\frac{2 i \pi }{k}
\mathrm{Tr} \left( 
Y_A^\dagger Y^A \psi^{B \dagger} \psi_B
-\psi^{B \dagger} Y^A Y_A^\dagger \psi_B 
-2 Y_A^\dagger Y^B \psi^{A \dagger} \psi_B
+2 \psi^{B \dagger} Y^A Y_B^\dagger \psi_A  \right. \CR
\left. 
-\epsilon^{ABCD} Y_A^\dagger \psi_B Y_C^\dagger \psi_D
+\epsilon_{ABCD} Y^A \psi^{B \dagger} Y^C \psi^{D \dagger} 
\right),
\label{vferm}
\end{eqnarray}
where the convention of the spinors 
is similar as in \cite{Klebanov}, but slightly different.\footnote{
Indices of a spinor are raised, 
$\theta^\alpha=\epsilon^{\alpha\beta} \theta_\beta$,
and lowered, $\theta_\alpha=\epsilon_{\alpha\beta} \theta^\beta$,
with $\epsilon^{12}=-\epsilon_{12}=1$.
The dirac matrix 
$(\gamma^\mu)_\alpha^{\ \beta}$ 
is taken such that $(\gamma^\mu)_{\alpha \beta}
\equiv (\gamma^\mu)_\alpha^{\ \gamma} \epsilon_{\beta \gamma}$ 
is a real symmetric matrics. 
We will use the gamma matrices with 
the first one lower and second one upper indices,
$(\gamma^\mu)_\alpha^{\ \beta}$,
only if the indices are surpressed.
The product of the two spinors 
are defined as $\theta \psi \equiv \theta^\alpha \psi_\alpha$
and 
$\theta \gamma^\mu \psi \equiv 
\theta^\alpha (\gamma^\mu)_\alpha^{\ \beta} \psi_\beta$
where we suppress the indices.
Note that 
$\theta^\alpha (\gamma^\mu)_{\alpha \beta} \psi^\beta
=-\theta^\alpha (\gamma^\mu)_\alpha^{\ \beta} \psi_\beta$
}
%and $k=8 \pi K$.

The ${\cal N}=6$ SUSY transformation 
is given by
\begin{eqnarray}
\delta Y^A &=& i \omega^{AB} \psi_B, \CR
\delta Y_A^\dagger &=& i \psi^{\dagger \, B} \omega_{AB}, \CR
\delta \psi_A &=& - \gamma_\mu \omega_{AB} D_\mu Y^B
+\frac{2 \pi}{k} \left( 
-\omega_{AB} (Y^C Y_C^{\dagger} Y^B-Y^B Y_C^{\dagger} Y^C)
+2 \omega_{CD} Y^C Y_A^\dagger Y^D
\right), \CR
\delta \psi^{A \dagger} &=& 
D_\mu Y_B^\dagger \omega^{AB} \gamma_\mu 
+\frac{2 \pi}{k} \left( 
-(Y_B^\dagger Y^C Y_C^\dagger -Y_C^\dagger Y^C Y_B^\dagger) \omega^{AB} 
+2 Y_D^\dagger Y^A Y_C^\dagger \omega^{CD} 
\right), \CR
\delta A_\mu &=& \frac{ \pi}{k} (
-Y^A \psi^{B \dagger} \gamma_\mu \omega_{AB} 
+ \omega^{AB} \gamma_\mu \psi_A Y_B^\dagger), \CR
\delta \hat{A}_\mu &=& \frac{ \pi}{k} (
-\psi^{A \dagger} Y^B \gamma_\mu \omega_{AB} 
+ \omega^{AB} \gamma_\mu Y_A^\dagger \psi_B), 
\label{susy}
\end{eqnarray}
where we assume that
$\psi$ and $\omega_{AB}$ have lower spinor indices,
while 
$\psi^\dagger$ and $\omega^{AB}$ have upper spinor indices,
even when the indices are suppressed and contracted.

By the 6 majorana (2+1)-dimensional spinors,
$\epsilon_i$ $(i=1,\ldots,6)$, which 
are the ${\cal N}=6$ SUSY generators, 
the $\omega_{AB}$ is given by 
\begin{eqnarray}
\omega_{AB} &=& \epsilon_i (\Gamma^i)_{AB}, \\
\omega^{AB} &=& \epsilon_i ((\Gamma^i)^*)^{AB},
\end{eqnarray}
in which the $A,B$ indices are anti-symmetric
and we take 4 by 4 matrices $\Gamma^i$ as follows:
\begin{eqnarray}
\Gamma^1 &=& \sigma_2 \otimes 1_2,
\;\;\; \Gamma^4 = -\sigma_1 \otimes \sigma_2, \CR
\Gamma^2 &=& -i \sigma_2 \otimes \sigma_3,
\;\;\; \Gamma^5 = \sigma_3 \otimes \sigma_2, \CR
\Gamma^3 &=& i \sigma_2 \otimes \sigma_1,
\;\;\; \Gamma^6 = -i 1_2 \otimes \sigma_2, 
\end{eqnarray}
which are chiral decomposed 
6-dimensional $\Gamma$-matrices.
These matrices satisfy
\begin{eqnarray}
&& \{ \Gamma^i, \Gamma^{j \dagger} \} = 2 \delta_{ij},
( \Gamma^i )_{AB} = - (\Gamma^i)_{AB}, \\ 
&& \frac{1}{2} \E^{ABCD} \Gamma^i_{CD}= 
-(\Gamma^{i \dagger})^{AB}=
((\Gamma^i)^*)^{AB}.
\end{eqnarray}
Therefore 
we have following relations
\begin{eqnarray}
(\omega^{AB})_\alpha &=& ((\omega_{AB})^*)_\alpha,
\;\;\;\;
\omega^{AB} =\frac{1}{2} \epsilon^{ABCD} \omega_{CD}.
\end{eqnarray}

We can explicitly check that 
the action (\ref{action}) is 
indeed invariant under the transformation 
(\ref{susy}).\footnote{
We can use the explicit representation 
of the gamma matrices as same as \cite{Klebanov},
i.e. $(\gamma^\mu)_\alpha^{\ \beta}=(i \sigma^2, \sigma^1,\sigma^3)$
and $(\gamma^\mu)_{\alpha \beta}=(-1, -\sigma^3,\sigma^1)$.
Another choice is $\gamma^\mu \rightarrow -\gamma^\mu$.
A parity transofrmation, 
$x^\mu \rightarrow -x^\mu, \, A_\mu \rightarrow -A_\mu$,
will change the overall sign of the Chern-Simons term
and the sign of the kinetic term of the fermions.
A charge conjugation, which interchanges
$(\Psi_A, Y^A, A_\mu)$ and 
$({\Psi^\dagger}^A, {Y^\dagger}_A, \hat{A}_\mu)$,
will change the overall sign of the Chern-Simons term
and replace $V_{ferm}$ to $-V_{ferm}$.
Thus, the actions with different signs of the $V_{ferm}$
are related by the two succecive transofrmations with
the gamma matrices which are given by 
$\gamma^\mu \rightarrow -\gamma^\mu$.
%If we consider the action given in \cite{Klebanov} in which
Moreover, 
the signs of the last two terms in (\ref{vferm}) are changed,
%it still have ${\cal N}=6$ SUSY
if we replace  
$\Gamma^i \rightarrow R \Gamma^i R$ where
$R=\left(
\begin{array}{cccc} 1 & 0 & 0 &0  \\ 0 & 1 & 0 &0  \\ 
0 & 0 & 0 &1   \\ 0 & 0 &  1 & 0 \end{array} 
\right)$.
Then the gamma matrices 
satisfy $ \frac{1}{2} \E^{ABCD} \Gamma^i_{CD}= 
(\Gamma^{i \dagger})^{AB}=
-((\Gamma^i)^*)^{AB}$.
}
We can also check that
if we restrict $\omega_{a \dot{b}}=0$, 
$(a=1,2, \dot{b}=3,4)$,
this transformation is same as the usual 
SUSY transformation of  
the ${\cal N}=2$ superfield formalism \cite{Klebanov}.
Note that since the superfield is written 
in the Wess-Zumino gauge,
the SUSY transformation is corrected by 
the super gauge transformation
with the gauge parameter proportional to  $\sigma$
and $\tilde{\sigma}$.
Including these, (\ref{susy}) will coincides with
the usual supersymmetry transformation in 
the superspace.

\section{M5-brane from the M2-brane action}

We consider solutions of the BPS equation of 
the ABJM action which corresponds to
the M2-branes ending on the M5-branes 
as in Basu-Harvey equation \cite{BaHa}. 
The BPS condition is $\delta \psi_A=0$. 
Here we will assume $Y^3=Y^4=0$ 
and $Y^1=Y^1(x^2), \; Y^2=Y^2(x^2)$,
namely the world-volume of the M5-branes
are along $\{x^0,x^1, x^4, x^5, x^6, x^7 \}$.  
We also assume
\begin{equation}
\gamma^2 \omega_{12} =\omega_{12}, \,\,
\gamma^2 \omega_{34} =\omega_{34},
\;\; \gamma^2 \omega_{a \dot{b}}=-\omega_{a \dot{b}},
\;\; \gamma^2 \omega_{\dot{b} a}=-\omega_{\dot{b} a},
\end{equation}
where 
$a=1,2$ and $\dot{b}=3,4$. 
Note that, for example, $\omega_{12}$ 
is a complex conjugate of $\omega_{34}$.
This means that we are considering 
a $\frac{1}{2}$ BPS solution, i.e. 
a solution with unbroken 6 supersymmetries.
We expect this will be obtained 
from the M5-M2-brane on ${\bf R}^{10,1}$,
which have unbroken 8 supersymmetries,
by the $\z_k$ orbifolding.

Then the SUSY transformation (\ref{susy}) for $\psi$
becomes
\begin{eqnarray}
0 &=&  
\frac{d Y^1}{d x^2} +\frac{2\pi}{k} (Y^2 Y_2^\dagger Y^1
-Y^1 Y_2^\dagger Y^2), \\ 
0 &=&
\frac{d Y^2}{d x^2} +\frac{2\pi}{k} (Y^1 Y_1^\dagger Y^2
-Y^2 Y_1^\dagger Y^1), 
\label{BPS} 
\end{eqnarray}
which can be written as
\begin{equation}
\frac{d Y^a}{d x^2} =-\frac{2\pi}{k} (Y^b Y_b^\dagger Y^a
-Y^a Y_b^\dagger Y^b).
\end{equation}
These equations have global $U(2)$ invariance which acts on $a,b$
indices
and $U(N) \times U(N)$ gauge invariance.

As in \cite{BaHa}, 
if we have $N \times N$ matrices $S^a$
which satisfy
\begin{eqnarray}
S^1 &=& S^2 S^{2 \dagger} S^1 -S^1 S^{2 \dagger} S^2  \CR
S^2 &=& S^1 S^{1 \dagger} S^2 -S^2 S^{1 \dagger} S^1,
\label{matS}
\end{eqnarray}
then 
\begin{equation}
Y^a= \sqrt{ \frac{k}{4 \pi x^2} } S^a, 
\label{s1}
\end{equation} 
$(x^2 >0) $ is the BPS solution represents
$N$ M2-brane ending on a M5-brane.\footnote{
$Y^a= \sqrt{ \frac{16 \pi k}{-x^2} } S^a$ $(x^2<0)$
with
$S^1 = S^1 S^{2 \dagger} S^2-S^2 S^{2 \dagger} S^1$ and
$S^2 = S^2 S^{1 \dagger} S^1 -S^1 S^{1 \dagger} S^2$ is 
also a BPS solution and 
represents an anti-M5-brane.
}
Instead of (\ref{s1}), 
\begin{equation}
Y^a=f^a(x^2) S^a,
\end{equation}
with
\begin{equation}
\frac{d f^1}{d x^2} + \frac{1}{2} |f^2|^2 f^1,
\;\; \frac{d f^2}{d x^2} + \frac{1}{2} |f^1|^2 f^2,
\end{equation}
is also a solution, which has 
a non-trivial real modulus,
We can assume without loss of generality 
that $f^i$ are real. 
Then, 
$C_0 \equiv |f^1|^2-|f^2|^2$ is a constant
and we obtain
\begin{equation}
 \frac{d (f^2)^2}{d x^2} 
+ \frac{1}{4} (f^2)^2 ((f^2)^2 +C_0) =0,
\end{equation}
which has a solution modulo the translation. 

For $N=2$, we have the following explicit solution
of (\ref{matS}),
\begin{eqnarray}
S^1 &=& \frac{1}{2} \left( 
\sigma_1+i \sigma_2 \right), \CR
S^2 &=& \frac{1}{2} \left( 
1_2 - \sigma_3 \right).
\end{eqnarray}
This solution seems strange as a fuzzy 3-sphere 
because $S^2$ is Hermite and diagonalized matrix,
thus it might not repersent an object extends in three directions.\footnote{
We thank S.~Kawai and S.~Sasaki for discussing this point.}
However, we note that $S^a$ is in a bi-fundamental representation,
instead of an adjoint representation and
there are $U(N) \times U(N)$ gauge symmetry,
instead of $U(N)$.
Therefore, we can always diagonalize $S^2$ and 
this solution may represent a fuzzy 3-sphere.
For arbitrary $N$, by the $U(N) \times U(N)$ gauge symmetry, 
we can take 
\beq
(S^2)_{i j } = \alpha_i \delta_{ij},
\eeq
where $\alpha^i$ is real and non-negative number.
We can further assume $\alpha_{i+1} \leq \alpha_{i}$
without loss of generality.
Then, from the first equation of (\ref{matS}),
we see that $(S^1)_{ij}=0$ if $(\alpha_i)^2-(\alpha_j)^2=1$.
This implies $S^1$ is block diagonalized
if $(\alpha_{i+1})^2=(\alpha_i)^2 -1$ is not satisfied 
for any $i=1, \cdots, N-1$.
The block diagonalized $S^1$ will represent several M5-branes.
Thus, we assume $(\alpha_{i+1})^2=(\alpha_i)^2 -1$, then
\beq
(S^1)_{ij}= \beta_{i} \, \delta_{i, \, j-1}  \,\,\, (i,j=1, \cdots, N).
\eeq
If we set $\beta_N=0$ and $\beta_0=0$ for convenience,
we can write $(S^1 (S^1)^\dagger )_{ij}= \delta_{ij} (\beta_{i})^2$
and $( (S^1)^\dagger S^1)_{ij}= \delta_{ij} (\beta_{i-1})^2$.
Now we can easily solve the second equation of (\ref{matS}),
\beq
( (\beta_{i})^2 - (\beta_{i-1})^2) \alpha_i= \alpha_i,
\,\,\, (i=1, \cdots N).
\eeq
Indeed, this implies that $\alpha_N =0$ for $i=N$ and 
$\beta_1=1$ for $i=1$.
(Here we have assumed $S^1$ is not block diagonalized. )
Therefore, we find
the BPS solution representing the $N$ M2-branes ending on a M5-brane
is (\ref{s1}) with 
\beq
(S^1)_{ij}= \delta_{i, \, j-1} \sqrt{i}, \,\,\, 
(S^2)_{i j } = \delta_{ij}  \sqrt{N-i} \, \,\,\,
(i,j=1, \cdots, N).
\label{solN}
\eeq
Of course, a diagonal sum of (\ref{solN}) is also a BPS
solution.\footnote{
This solution was obtained also in \cite{Go}.}
%We expect that there exist solutions for $N>2$
%and represent fuzzy 3-spheres. 

We can estimate the tension of the M5-brane.
In the large $N$ limit,
the approximate radius of the fuzzy 3-sphere is $r \sim \sqrt{ k
N/ (4 \pi x^2)}$.
The action is evaluated as
\beq
S \sim -2 \int d^3 x   \mathrm{Tr} D_{\mu} Y_a^{\dagger} D^{\mu} Y^a
\sim -2 \int d^3 x \frac{ k}{16 \pi  (x^2)^3}  \mathrm{Tr} (S^a (S^a)^\dagger)
\sim - \int dx^0 dx^1 dr r^3  \frac{2 \pi}{ k },
\eeq
and the area of the three dimensional sphere $2 \pi^2$ 
should be divided by $k$
because of the $\z_k$ orbifolding.
Thus, the tension of the M5-brane is independent of $k$ and $N$ as expected.

For the fuzzy 2-sphere in D1-branes ending on D3-branes,
we can obtain the non-commutative $R^2$ 
by taking a limit which corresponds to
focusing on the north pole of the fuzzy 2-sphere.  
We will consider a similar limit for our 
fuzzy 3-sphere. 
The equations (\ref{matS}) can be written
by four Hermite matrices as
\begin{eqnarray}
A &=&  i \left(
[B,C^2+D^2] +\{ A,[C,D] \}  \right), \CR
B &=&  i \left(
-[A,C^2+D^2] +\{ B,[C,D] \}  \right), \CR
C &=&  i \left(
[D,A^2+B^2] +\{ C,[A,B] \}  \right), \CR
D &=&  i \left(
-[C,A^2+B^2] +\{ D,[A,B] \}  \right), 
\label{eq4}
\end{eqnarray}
where 
\begin{equation}
S^1=A+iB, \;\;\; S^2=C+iD.
\end{equation}
We assume $A=\Lambda+\delta$,
where $\Lambda \gg 1$ is a constant,
and $B= 0$.
%$[\delta,C] \sim 0$, $[\delta,D] \sim 0$.
Then (\ref{eq4}) becomes
\begin{eqnarray}
[C,D] = -\frac{i}{2}, \;\;\; \; 
C=i [D, 2 \Lambda \delta], \;\; D=-i [C,2 \Lambda \delta], \,\,
[\delta, C^2+D^2]=0, 
\end{eqnarray}
which can be solved as
\begin{equation}
C=\frac{1}{\sqrt{2}} \hat{p}, \;\;
D= \frac{1}{\sqrt{2}} \hat{q}, \;\;
- 4 \Lambda \delta=\hat{p}^2+ \hat{q}^2 +const.
\end{equation}
In the limit which take the M2-branes to 
D2-branes %\footnote{
%In Appendix A, we will explicitly show that
%the action (\ref{action}) becomes 
%the (2+1)d maximal supersymmetric Yang-Mills action
%in the limit.}
\cite{M2toD2, Mfold, ABJM},
$B$ is the compactified direction and 
$C$ and $D$ span the non-commutative 2-plane.

\section{M5-branes with flux}

The M5-brane with flux can be considered as the 
bound state of M2-branes and M5-branes.
We expect that there are
solitonic solutions in the action (\ref{action})
which represent the bound states.
Because the M5-brane extending 
in $\{x^0,x^1,x^2 \}$ and three directions in $C^4/\z_k$,
the supersymmetries will be completely broken. 
Actually, the action does not have 
additional non-linearly realized supersymmetry which 
would restore supersymmetry. 
Therefore, we will study the equations of motion,
instead of BPS equations.

First, by the $2 N \times 2N$ Hermitian matrices 
\begin{equation}
\tilde{Y}_A=\matt[0,Y^A,Y_A^\dagger ,0],
\end{equation}
the bosonic potential can be written in a simple form
\begin{equation}
V_{bos} \sim {\rm Tr} [
(\tilde{Y}_A (\tilde{Y}_B \tilde{Y}_B) 
-(\tilde{Y}_B \tilde{Y}_B) \tilde{Y}_A)^2 
-2 (\tilde{Y}_A \tilde{Y}_B \tilde{Y}_C
-\tilde{Y}_C \tilde{Y}_B \tilde{Y}_A)^2 ]. 
\label{bos}
\end{equation}
Now we assume $Y^A$ are constant Hermite
matrices.
We further assume that 
\begin{equation}
\alpha_{A \; C}^{\;\; B}
\equiv Y_{A}^{ \dagger} Y^{B} 
Y_{C}^{\dagger} -Y_{C}^{\dagger} Y^{B} Y_{A}^{\dagger},
\label{M5a}
\end{equation}
is proportional to the $N \times N$ unit matrix,
$1_N$, thus they commute with any field. 
Note that $\alpha_{A \; C}^{\;\; B}$ is 
an anti-Hermitian
and anti-symmetric under exchange of the indices $A$ and $C$.
Then, we can see from (\ref{bos}) that 
the equations of motion are solved 
if 
\begin{equation}
\alpha_{A \; C}^{\;\; B}+
\alpha_{C \; B}^{\;\; A} + \alpha_{B \; A}^{\;\; C}=0,
\label{M5}
\end{equation} 
is satisfied.
We set $Y^4=0$, then $A,B,C$ runs $1$ from $3$ and 
the configurations (\ref{M5a}) with
(\ref{M5}) may represent a bound state of a M5-brane and
M2-branes. 
Note that 
by taking the trace of (\ref{M5a}) and using 
the relation (\ref{M5}),
we can see that
the configurations (\ref{M5a}) 
can not be realized if $N$ is finite, thus
we need infinitely many M2-branes,
like the D4-D2 bound state in the D2-brane picture.

Because of (\ref{M5}),
there are 8 independent components of 
$\alpha_{A \; C}^{\;\; B}$.
These should correspond to the flux on 
the M5-brane, if there are indeed 
M5-brane solutions
for (\ref{M5a}) and (\ref{M5}).
It is very important to find 
explicit solutions of (\ref{M5a}) and (\ref{M5})
in order to establish these indeed represent 
the bound state.

\section{Conclusions and discussion}

In this paper, we have studied 
the BPS equations 
of the ABJM action, which will describe 
the M5-brane.
We have found solutions of these equations.
These BPS equations are
analogues of the Basu-Harvey equation
\cite{BaHa} and
we expect that 
the solutions represent
$N$ M2-branes ending on the M5-brane.
We also discussed the M5-M2 bound state
as solutions of the equations of motion,
instead of the BPS equation.
It is very interesting to
investigate the properties of 
the M5-branes by the solutions.

We can easily extend our study in this paper 
to some modifications of the ABJM actions,
for example, to the orbifold theories 
\cite{Klebanov, TeYa, Fuji}.

For the Nahm equation and their string theory 
realization \cite{Do, Di},
we have an $\alpha'$ exact equivalence 
between the D2-brane picture (Nahm equation)
and the D4-brane picture (Monopole equation)
\cite{HaTe3} using the tachyon condensation \cite{Te1}.
It is interesting to
see how these results are lifted to the M2-brane case. 

% We can also consider the 

% Of course, our discussion in this paper is 
% rather naive and need further study.
% In particular the problem of the unitarity may be important.

% To make this clear is an interesting question.

% Another question is how to 

% Since it mixes the coordinate and the fields in general,
% to find the unbroken symmetry will be interesting.

\vskip6mm
\noindent
{\bf Acknowledgements}

\vskip2mm
We would like to thank S. Kawai, T. Takayanagi and F. Yagi
for useful discussions.
S.~T.~is partly supported by the Japan Ministry of Education, Culture, Sports, Science and Technology.
\\

\noindent
{\bf Note added in proof}: 

As this article was being completed,
we received the preprints \cite{Gaiotto, Hosomichi3}
which also present the ${\cal N}=6$ supersymmetry
transformation, in different forms.

\appendix
\setcounter{equation}{0}

% \section{(2+1)d Yang-Mills limit}

\end{document}